\documentclass[aps,prd,twocolumn,amsmath,superscriptaddress,amssymb,showpacs,floatfix,nofootinbib]{revtex4-1}

\usepackage{graphicx}
\usepackage{dcolumn}
\usepackage{xcolor}
\usepackage{bm}
\usepackage{natbib}
\usepackage{calligra}
\usepackage[T1]{fontenc}
\usepackage{egothic}
\usepackage[T1]{fontenc}
\newfont{\rsfsten}{rsfs10 scaled 1200}
\newfont{\rsfsseven}{rsfs10 scaled 1200}
\newfont{\rsfsfive}{rsfs10 scaled 1200}
\usepackage{epsfig}
\usepackage{units}
\usepackage[utf8]{inputenc}

\newcommand{\be}{\begin{equation}}
\newcommand{\ee}{\end{equation}}
\newcommand{\bea}{\begin{eqnarray}}
\newcommand{\eea}{\end{eqnarray}}

\newcommand{\ie}{{\it i.e.~}}

\def\lsim{\mathrel{\raise.3ex\hbox{$<$\kern-.75em\lower1ex\hbox{$\sim$}}}}
\def\gsim{\mathrel{\raise.3ex\hbox{$>$\kern-.75em\lower1ex\hbox{$\sim$}}}}

\begin{document}

\hspace{13cm} \parbox{5cm}{FERMILAB-PUB-17-010-A}

\hspace{13cm}
\vspace{0.6cm}

\title{Possible Evidence for the Stochastic Acceleration of \\ Secondary Antiprotons by Supernova Remnants}

\author{Ilias Cholis}
\email{icholis1@jhu.edu, ORCID: orcid.org/0000-0002-3805-6478}
\affiliation{Department of Physics and Astronomy, The Johns Hopkins University, Baltimore, Maryland, 21218, USA}
\author{Dan Hooper}
\email{dhooper@fnal.gov, ORCID: orcid.org/0000-0001-8837-4127}
\affiliation{Fermi National Accelerator Laboratory, Center for Particle Astrophysics, Batavia, Illinois, 60510, USA}
\affiliation{University of Chicago, Department of Astronomy and Astrophysics, Chicago, Illinois, 60637, USA}
\affiliation{University of Chicago, Kavli Institute for Cosmological Physics, Chicago, IL 60637, USA}
\author{Tim Linden}
\email{linden.70@osu.edu, ORCID: orcid.org/0000-0001-9888-0971}
\affiliation{Center for Cosmology and AstroParticle Physics (CCAPP) and Department of Physics, The Ohio State University Columbus, OH, 43210 }

\date{\today}

\begin{abstract}

The antiproton-to-proton ratio in the cosmic-ray spectrum is a sensitive probe of new physics. Using recent measurements of the cosmic-ray antiproton and proton fluxes in the energy range of 1\,--\,1000~GeV, we study the contribution to the $\bar{p}/p$ ratio from secondary antiprotons that are produced and subsequently accelerated within individual supernova remnants.  We consider several well-motivated models for cosmic-ray propagation in the interstellar medium and marginalize our results over the uncertainties related to the antiproton production cross section and the time-, charge-, and energy-dependent effects of solar modulation. We find that the increase in the $\bar{p}/p$ ratio observed at rigidities above $\sim$\,100~GV cannot be accounted for within the context of conventional cosmic-ray propagation models, but is consistent with scenarios in which cosmic-ray antiprotons are produced and subsequently accelerated by shocks within a given supernova remnant.  In light of this, the acceleration of secondary cosmic rays in supernova remnants is predicted to substantially contribute to the cosmic-ray positron spectrum, accounting for a significant fraction of the observed positron excess.

\end{abstract}


\maketitle


The ratio of cosmic-ray (CR) antimatter to matter is a powerful probe of new physics, in particular of dark matter annihilation or decay~\cite{Bergstrom:1999jc, Hooper:2003ad, Bertone:2004pz, Profumo:2004ty, Bringmann:2006im}. However, antimatter can also be produced astrophysically through the interactions of CR protons with gas. As the astrophysical flux of CR antimatter depends on CR propagation, its measurement depends on diffusion in the interstellar medium (ISM)~\cite{Pato:2010ih,Simet:2009ne,DiBernardo:2009ku,Strong:2007nh}. Exotic contributions to the antimatter flux can be differentiated from conventional astrophysics through the observation of secondary CRs, such as boron, that are produced via spallations, but not from dark matter.

Over the past decade, an intriguing rise with energy in the CR positron fraction ($e^+/(e^++e^-)$) has been observed by both \textit{PAMELA} \cite{Adriani:2010rc} and \textit{AMS-02}~\cite{Aguilar:2013qda}. The dark matter interpretation of this excess has received significant attention~\citep{Bergstrom:2008gr, Cirelli:2008jk, Cholis:2008hb,  Cirelli:2008pk, Nelson:2008hj, ArkaniHamed:2008qn, Cholis:2008qq, Cholis:2008wq, Harnik:2008uu, Fox:2008kb, Pospelov:2008jd, MarchRussell:2008tu, Chang:2011xn}. Dark matter models that explain the positron fraction typically invoke particles with masses of $\sim$\,1\,--\,3~TeV annihilating or decaying to e$^+$e$^-$ pairs through intermediate two- or three-body decays~\citep{Cholis:2013psa, Cirelli:2008pk, Dienes:2013xff, Finkbeiner:2007kk}. Alternatively, the positron excess could very plausibly be generated by nearby pulsars, with ages of $\sim$\,$10^{5}$\,--\,$10^{6}$ years~\cite{Hooper:2008kg, Yuksel:2008rf, Profumo:2008ms, Malyshev:2009tw, Grasso:2009ma, Linden:2013mqa,Cholis:2013psa}.

A third explanation for the rising positron fraction includes a two-step process where positrons are first produced via hadronic interactions (followed by pion/muon decay) within supernova remnants (SNRs), and are then accelerated by shocks within those same remnants before escaping into the ISM~\cite{Blasi:2009hv, Mertsch:2009ph, Ahlers:2009ae, Blasi:2009bd, Mertsch:2014poa}. In contrast to dark matter or pulsar scenarios, this ``stochastic acceleration'' predicts a similar rise for \emph{all} species of CR secondaries produced via hadronic interactions (see also \cite{Fujita:2009wk, Kohri:2015mga, Malkov:2016kbe}). In Ref.~\cite{Cholis:2013lwa}, the observed boron-to-carbon (B/C) ratio was utilized to show that stochastic acceleration could not account for the entirety of the positron excess, though this assumes that both CR protons and carbon nuclei are produced equally across the population of SNRs (see, however,~\cite{Kachelriess:2011qv, Kachelriess:2012ag}).

\begin{table*}[t]
    \begin{tabular}{ccccccccc}
         \hline
           Model \,\,&\,\, $\delta$ \,\,&\,\, $z_{L} (kpc)$ \,\,&\,\, $D_{0} \times 10^{28}$ (cm$^2$/s) \,\,&\,\, $v_{A}$ (km/s) \,\,&\,\, $dv_{c}/dz$ (km/s/kpc) \,\,&\,\, $\alpha_{1}$ \,\,&\,\, $\alpha_{2}$ \,\,&\,\, $R_{\rm br}$ (GV)\\
            \hline \hline
            C &  0.40 & 5.6 & 4.85 & 24.0 & 1.0 & 1.88 & 2.38 & 11.7 \\      
            E &  0.50 & 6.0 & 3.10 & 23.0 & 9.0 & 1.88 & 2.45 & 11.7 \\
            F &  0.40 & 3.0 & 2.67 & 22.0 & 3.0 & 1.87 & 2.41 & 11.7 \\
            \hline \hline 
        \end{tabular}
\caption{Key parameters for the models used to describe the injection and propagation of cosmic rays in the Galaxy. Assuming isotropic and homogeneous diffusion, the diffusion tensor simplifies to a coefficient $D_{xx}(R) = D_{0} (R/4 GV)^{\delta}$ within a zone of half-height, $z_{L}$, where $R \equiv p/|q|$ is the absolute value of the cosmic-ray rigidity. $v_{A}$ is the Alfv$\acute{\textrm{e}}$n speed and $dv_{c}/dz$ the convection speed gradient perpendicular to the Galactic Disk. CR protons are injected with a differential spectrum of $dN_{p}/dR \propto R^{-\alpha}$, where $\alpha_{1}$ and $\alpha_{2}$ are the spectral indices below and above $R_{\rm br}$.} 
\label{tab:ISMBack}
\end{table*}

These constraints are mitigated if individual SNRs produce varying relative abundances of different primary CR species. For example, if nearby SNRs are efficient accelerators of secondaries, but have low abundances of intermediate mass nuclei, then the connection between the B/C ratio and the positron fraction could be weakened~\cite{Cholis:2013lwa}. However, a direct comparison exists between the positron fraction and the antiproton-to-proton ratio ($\bar{p}/p$), since secondary antiprotons and positrons are both generated through proton-proton interactions. In this paper, we examine $\bar{p}/p$ in stochastic acceleration models. We find evidence for an excess of high-energy antiprotons measured with great accuracy by \cite{Aguilar:2016kjl}, that can be explained by stochastic acceleration and that cannot be accounted for by uncertainties in solar modulation, cosmic-ray propagation or the antiproton production cross section. Intriguingly, our results suggest that a significant fraction of the observed positron excess originates from the secondary acceleration of positrons in SNRs.



There are a number of systematic uncertainties that must be treated carefully to interpret the $\bar{p}/p$ ratio measured by \textit{AMS-02}. In particular, we consider uncertainties associated with CR propagation in the ISM, the antiproton production cross section, and the effects of solar modulation. 

\begin{figure}
\hspace{-0.23in}
\includegraphics[width=3.60in,angle=0]{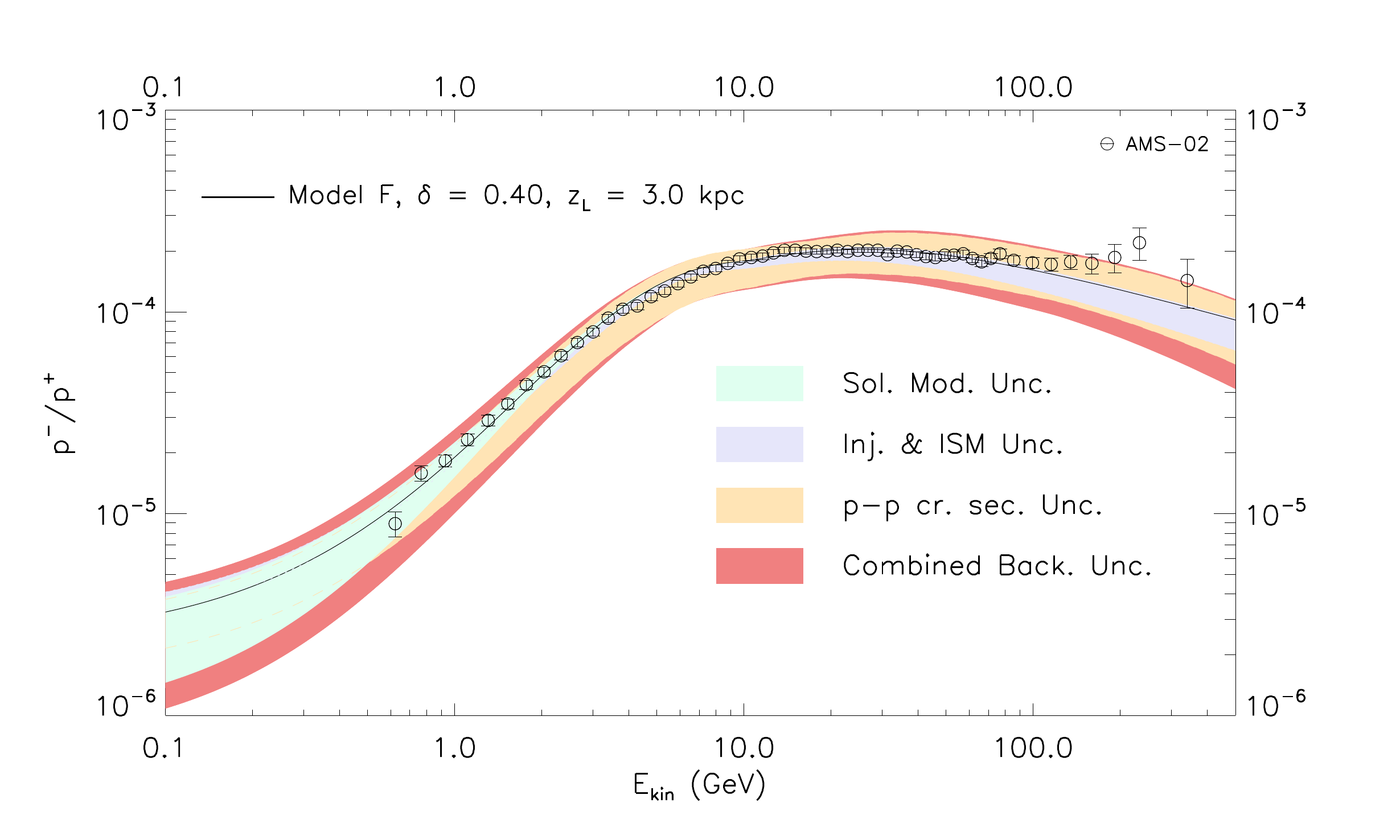}
\vskip -0.1in
\caption{The $\bar{p}/p$ ratio measured by \textit{AMS-02}, compared to predictions from conventional secondary production (without accelerated secondaries). Results are shown for model F using central values for the antiproton production cross section and solar modulation parameters (solid black line). Variations in the propagation model, antiproton production cross section, and solar modulation parameters produce the range of predictions shown as the blue, orange, and green bands, respectively. The combination of these uncertainties is represented by the red band.}
\label{fig:UncertBands}
\end{figure}

Most CR antiprotons are produced through the hadronic interactions of high-energy protons and nuclei with interstellar gas. To model the injection and propagation of CRs through the Galaxy, we utilize \texttt{Galprop}, which numerically solves the transport equation to calculate the local flux of primary and secondary CR species~\cite{GALPROPSite, Strong:2015zva, NEWGALPROP, galprop}. The primary uncertainties in this calculation are the injected spectrum and distribution of primary CRs, and the timescales for diffusion through the Galactic medium. Convection and diffusive re-acceleration can also be relevant. Numerous measurements provided by \textit{AMS-02}, \textit{PAMELA} and \textit{Voyager 1} constrain the characteristics of CR propagation. In this work, we follow the procedure described in Ref.~\cite{Cholis:2015gna} from which we take two propagation models (models C and E). We additionally produce a new (thin disk) model (model F). Each model provides a good fit to the proton spectrum measured by \textit{Voyager 1} and \textit{PAMELA}, and the B/C data from \textit{AMS-02} and  \textit{PAMELA}. We use these three models to envelope the uncertainties related to CR production and propagation.  
These models are summarized in Table~\ref{tab:ISMBack} (see Ref.~\cite{Cholis:2015gna}), and their predictions for the $\bar{p}/p$ ratio  are shown as a blue band in Fig.~\ref{fig:UncertBands} (labelled as ``Inj.~\& ISM Unc.''). We also show the $\bar{p}/p$ ratio predicted by model F (solid black line).


\begin{table*}[t]
\begin{tabular}{ccccc}
\hline
Era & \,\,\,\, $|B_{\rm tot}|$ (nT) \,\,\,\,& \,\,\,\,$\alpha$ (degrees) \,\,\,\,& \,\,\,\,$N'(q>0) \cdot H(-qA(t))$ \,\,\,\,&\,\,\,\,  $N'(q<0) \cdot H(-qA(t))$ \\
\hline \hline
07-12/11 & 4.7 & 60.5 & 1 & 0 \\
01-06/12 & 4.8 & 67.2 & 1 & 0 \\
07-12/12 & 5.3 & 70.0 & 0.67 & 0.33 \\
01-06/13 & 5.5 & 71.0 & 0.50 & 0.50 \\
07-12/13 & 5.2 & 70.0 & 0.33 & 0.67 \\
01-06/14 & 5.3 & 67.0 & 0 & 1 \\
07-12/14 & 5.6 & 62.0 & 0 & 1 \\
01-06/15 & 6.6 & 56.6 & 0 & 1 \\
\hline \hline
\end{tabular}
\caption{The values of $|B_{\rm tot}|$ and $\alpha$ as averaged over each six-month interval within the period of  \textit{AMS-02}  observations (May 2011- May 2015). We also list the values of $N'(q) \cdot H(-qA(t))$, as appearing in Eq.~\ref{eq:ModPot}, for both protons and antiprotons.}
\label{tab:AMS_HMF}
\end{table*}

We emphasize that the decreasing $\bar{p}/p$ at high energies is a generic feature of any leaky-box diffusion model. 
Since the diffusion coefficient has an energy dependence $D \propto E^{\delta}$, the
total grammage encountered by cosmic-ray primaries falls as $E^{\delta}$, and this softened spectrum is inherited by cosmic-ray secondaries. These secondaries are themselves softened by diffusive escape, leading to a primary-to-secondary ratio which falls as $E^{\delta}$. Even if as observed above 0.5 TeV the proton spectrum becomes harder by 0.1 in its power-law spectrum value, and with those protons collisions  giving the antiprotons at $\simeq$100 GeV in energy; a value of $\delta \simeq 0.5$ will still give a $\bar{p}/p$ ratio that falls with increasing energy. 
The nearly energy-independent
$\bar{p}/p$ ratio  observed by \textit{AMS-02} at energies above $\sim$100~GeV \cite{Aguilar:2016kjl},
can only be accommodated in models with $\delta$~$\simeq$~0, which are strongly ruled out~\cite{Trotta:2010mx,Pato:2010ih,Simet:2009ne,DiBernardo:2009ku,Strong:2007nh}. Thus, the error band shown in Fig.~\ref{fig:UncertBands} is generically applicable to any \texttt{Galprop} model consistent with observations.

The cross section for antiproton production in inelastic $p-p$ collisions has been carefully studied~\cite{Tan:1982nc, Tan:1983de, Duperray:2003bd}, using data from Refs.~\cite{Dekkers:1965zz, Capiluppi:1974rt, Allaby:1970jt, Guettler:1976ce, Johnson:1977qx, Antreasyan:1978cw}. However, there remain significant uncertainties reagrding the production for for antiprotons in the collisions of CR protons and nuclei. While \texttt{Galprop v54} handles the production of antiprotons~\cite{Moskalenko:2001ya}, it does not include the most recent measurements of the antiproton production cross section~\cite{Arsene:2007jd, Anticic:2009wd}, nor does it account for the uncertainties in this quantity, which can be significant in the determination of the $\bar{p}/p$ ratio~\cite{Bringmann:2014lpa, Hooper:2014ysa, Cuoco:2016eej}. 

Recently, several groups have studied and quantified the uncertainties in the antiproton production cross section~\cite{diMauro:2014zea, Kappl:2014hha, Kachelriess:2015wpa}. 
To fit the $\textit{AMS-02}$ $\bar{p}/p$ data \cite{Aguilar:2016kjl}, we first calculate the antiproton spectrum for a given propagation model, and then renormalize those fluxes by the following continuous (in kinetic energy, prior to solar modulation, $E_{\textrm{kin}}^{\textrm{ISM}}$) function:
\begin{eqnarray}
N_{CS}(E_{\textrm{kin}}^{\textrm{ISM}}) &=& a + b~\ln \bigg(\frac{E_{\textrm{kin}}^{\textrm{ISM}}}{{\rm GeV}}\bigg) + c \bigg[\ln\bigg(\frac{E_{\textrm{kin}}^{\textrm{ISM}}}{{\rm GeV}}\bigg)\bigg]^{2} \nonumber \\
&+& d \bigg[\ln\bigg(\frac{E_{\textrm{kin}}^{\textrm{ISM}}}{\rm{GeV}}\bigg)\bigg]^{3}.
\label{eq:CSpbar}
\end{eqnarray}
We bound the values of $a$, $b$, $c$ and $d$ such that $N_{CS}(E_{\textrm{kin}}^{\textrm{ISM}})$ resides within the 3$\sigma$ uncertainties presented in~\cite{diMauro:2014zea}. We add $10\%$  uncertainty in Eq.~\ref{eq:CSpbar} to account for the local galactic gas uncertainties. 
The impact of this uncertainty is shown in Fig.~\ref{fig:UncertBands}, by the 
orange band surrounding the central prediction of propagation model F (labeled ``p-p cr.~sec.~Unc.'').

As CRs enter the Solar System, they experience heliospheric forces resulting in solar modulation. In treating solar modulation, we adopt the standard formula:
\begin{eqnarray}
\frac{dN^{\oplus}}{dE_{\textrm{kin}}} (E_{\textrm{kin}}) &=& \frac{(E_{\textrm{kin}}+m)^{2} -m^{2}}{(E_{\textrm{kin}}+m+|Z| e \Phi)^{2} -m^2} \nonumber \\ 
&\times& \, \frac{dN^{\rm ISM}}{dE_{\rm kin}^{\rm ISM}} (E_{\rm kin}+| Z | e \Phi),
\label{eq:Mod}
\end{eqnarray}
where $E_{\textrm{kin}}$ is the kinetic energy of CRs at Earth, $|Z|e$ is their charge, $dN^{\oplus}/dE_{\textrm{kin}}$ the differential CR flux at Earth, and $dN^{\rm ISM}/dE_{\rm kin}^{\rm ISM}$ is the local ISM differential flux. $\Phi$ is the modulation potential, for which we use the predictive time-, charge- and rigidity-dependent formula presented in~\cite{Cholis:2015gna}:
\begin{eqnarray}
\label{eq:ModPot}
\Phi(R,t,q) &=& \phi_{0} \, \bigg( \frac{|B_{\rm tot}(t)|}{4\, {\rm nT}}\bigg) + \phi_{1} \, N'(q) H(-qA(t)) \\ 
&\times& \bigg( \frac{|B_{\rm tot}(t)|}{4\,  {\rm nT}}\bigg) \, \bigg(\frac{1+(R/R_0)^2}{\beta (R/R_{0})^3}\bigg) \, \bigg( \frac{\alpha(t)}{\pi/2} \bigg)^{4}, \nonumber
\end{eqnarray}
where $B_{\rm tot}(t)$ is the strength of the heliospheric magnetic field (HMF) measured at Earth, $A(t)$ is its polarity, and $\alpha(t)$ the tilt angle of the heliospheric current sheet.  $R$, is the CR rigidity before entering the heliosphere (see Refs.~\cite{Cholis:2015gna, Potgieter:2013pdj}). 
$N'(q)$ is $\neq 1$ during eras in which the HMF does not have a well-defined polarity. 
We adopt $R_{0} = 0.5$ GV and marginalize over the solar modulation uncertainties described in Ref.~\cite{Cholis:2015gna}, allowing $\phi_{0} \in [0.32,0.38]$ GV and $\phi_{1} \in [0,16]$ GV.

\begin{table*}[t]
    \begin{tabular}{cccccccc}
         \hline
           ISM  mod.& $ \phi_{0}$ & $\phi_{1}$ & a & b & c & d & $\chi^{2}_{\rm tot}$(/d.o.f.) \\
            \hline \hline
            C & 0.32 & 4.0 & 1.26 & -0.125 & -0.010 & 0.006 & 44.0 (0.86)  \\
            E & 0.32 & 9.2 & 0.83 & 0.170 & -0.046 & 0.007 & 59.6 (1.17) \\
            F & 0.32 & 15.6 & 0.94  & 0.055 & -0.032 & 0.006 & 58.4 (1.15) \\ 
            \hline \hline 
        \end{tabular}
       \caption{The best-fit parameters for propagation models C, E and F, assuming that the observed antiprotons are secondaries produced only in the ISM (\ie neglecting stochastic acceleration).} 
    \label{tab:fitBackBestFit}
\end{table*}

\begin{table*}[t]
    \begin{tabular}{ccccccc}
         \hline
           ISM mod. & $K_{B}^{\textrm{best}}$ & $K_{B}^{95\% \textrm{upper}}$ & $K_{B}^{95\% \textrm{lower}}$ & $\chi^{2}_{tot}$ & $\chi^{2}_{d.o.f.}$ & $\Delta\chi^{2}_{tot}$(from back only) \\
            \hline \hline
            C & 6.1 & 7.6 & 4.6 & 34.0 & 0.68 & 10.0   \\
            E & 10.4 & 12.4 & 8.1 & 39.9 & 0.80 & 19.7   \\
            F & 7.4 & 8.9 & 5.7 & 37.5 & 0.75 & 20.9   \\
            \hline \hline 
        \end{tabular}
       \caption{The best-fit value and 95\% confidence level upper and lower limits on $K_B$, for each propagation model. We show the $\chi^2$ fit to the $\bar{p}/p$ spectrum measured by \textit{AMS-02}, and the improvement to the fit relative to the case without acceleration of secondaries ($K_B=0$, of Table~\ref{tab:fitBackBestFit}). The fit consistently prefers positive values of $K_B$, at a level of $\Delta \chi^{2} \simeq$\,10.0\,--\,20.9, corresponding to a 3.2\,--\,4.6$\sigma$ preference for the acceleration of secondary antiprotons in SNRs.}
    \label{tab:fitStochAccBestFit}
\end{table*}

Given that the $\bar{p}/p$ data of \cite{Aguilar:2016kjl} utilized in this study, have been taken over several years, between May 2011 and May 2015, we must account for the time-evolving properties of the HMF. We note that the modulation potential, $\Phi$, depends on both $|B_{\rm tot}|$ and $\alpha$, and thus is not linear with time. To account for this, we break the model into six-month periods and use the time-averaged values of $|B_{\rm tot}|$ (from the \textit{ACE} magnetometer~\cite{ACESite}) and $\alpha$ (calculated by the Wilcox Solar Observatory~\cite{WSOSite}) for each interval (see Table~\ref{tab:AMS_HMF}). For periods where the HMF geometry was being re-configured, we adopt values of $N'(q) <1 $ chosen to result in a smooth transition of the second term in Eq.~\ref{eq:ModPot}. We use these values of $|B_{\rm tot}|$, $\alpha$ and $N'(q)$ to calculate the modulated spectra for each individual six-month period, and then combine these eras to determine the total CR spectrum over the period observed by \textit{AMS-02}. The impact of the uncertainties related to solar modulation is depicted in Fig.~\ref{fig:UncertBands} by the green band (labelled ``Sol.~Mod.~Unc.'').


To combine the uncertainties associated with propagation through the ISM, the antiproton production cross section, and solar modulation, we calculate our fit to the \textit{AMS-02} $\bar{p}/p$ data for each of our three propagation models, marginalizing over uncertainties in the parameters $\phi_{0}$, $\phi_{1}$, $a$, $b$, $c$, and $d$. The best-fit parameters for each propagation model are shown in Table~\ref{tab:fitBackBestFit},
while the range of the combined uncertainties is depicted by the red band in Fig.~\ref{fig:UncertBands}. At kinetic energies below 1 GeV, the largest source of uncertainty is solar modulation. Between 2\,--\,20 GeV the main uncertainty is the antiproton production cross section. Above $\sim$\,20 GeV, uncertainties in the antiproton production cross section and CR propagation are both important. 




We now consider the stochastic acceleration of CR secondaries in SNRs. We assume that SNR shocks are supersonic, with a compression 
ratio of $v^{-}/v^{+} = 4$, where $v^{+}$ is the plasma down-stream velocity and $v^{-}$ is the plasma up-stream velocity, both defined in the frame of the shock front. 
As particles are accelerated inside the SNR, to a spectrum $N_{j}$, they interact with the dense gas and spallate with a partial cross-section $\sigma^{sp}_{j \rightarrow i}$, to produce lighter species $i$, or decay to them with a time-scale of  $\tau^{dec}_{j \rightarrow i}$~\cite{Blasi:2009hv, Mertsch:2009ph, Cholis:2013lwa}. 
For these lighter species, the source term is: 
\begin{equation}
Q_{i}(E_{kin}) = \Sigma_{j}N_{j}(E_{kin}) \left[ \sigma^{sp}_{j \rightarrow i} \, \beta \, c \, n_{gas} + \frac{1}
{\frac{E_{kin}}{1 \, GeV}\,\tau^{dec}_{j \rightarrow i}} \right].
\label{eq:sourceTerm}
\end{equation}
$n_{gas}$ is the gas density where the spallations occur and $E_{kin}$ is the kinetic energy per nucleon. 

These secondaries then undergo further spallations and decays at a rate:
\begin{equation}
\Gamma_{i}(E_{\rm kin}) =  \sigma^{\rm sp}_{i} \, \beta \, c \, n_{\rm gas} + \frac{1}{\frac{E_{\rm kin}}{1 \, GeV}\,\tau^{\rm dec}_{i}}, 
\label{eq:CRLossRate}
\end{equation}
where  $\sigma^{\rm sp}_{i}$ and $\tau^{\rm dec}_{i}$ are the spallation cross section and decay lifetime of nuclei species, $i$, respectively. 
Including to the above, advection, diffusion, and adiabatic energy losses, one gets the transport equation for species $i$:
\begin{equation}
v \frac{\partial f_{i}}{\partial x} = D_{i}\frac{\partial^{2} f_{i}}{\partial x^{2}} + \frac{1}{3}\frac{d v}{d x}p\frac{\partial f_{i}}{\partial p} - \Gamma_{i}f_{i} + q_{i}.
\label{eq:TranspEq}
\end{equation}
$D_{i}$ is the diffusion coefficient, $v$ the advection velocity, $f_{i}$ the phase space density of species $i$ and $q_{i}$ the relevant source term.

If enough CRs of species $i$ are produced and accelerated in the SNR before spallating or decaying (1/$\Gamma_{i}$ $\gg$ $\tau^{acc}$), they can have a significant impact on the observed secondary-to-primary ratios. Following Refs.~\cite{Blasi:2009hv, Mertsch:2009ph, Cholis:2013lwa}, we assume Bohm diffusion for CRs around the shock front:
\begin{eqnarray}
D_{i}^{\pm}(E) &=& \frac{K_{B}\, r_{L}(E)\,c}{3} \\
&=& 3.3 \times 10^{22} \, {\rm cm}^{2} \,{\rm s}^{-1} \times K_{B} \, \bigg(\frac{\mu{\rm G}}{B}\bigg)\, \bigg(\frac{E}{\rm GeV}\bigg) \, \bigg(\frac{1}{Z_i}\bigg), \nonumber
\label{eq:BohmDiff}
\end{eqnarray}
where $r_{L}$ is the Larmor radius, $B$ is the magnetic field, and $Z$ and $E$ are the charge and energy of the CR. $K_{B}$ is a factor~\cite{Mertsch:2009ph} scaling as $K_{B} \simeq (B/\delta B )^{2}$ \cite{Blasi:2009hv}, allowing for faster diffusion of CRs around the shock front. Measurements of the B/C ratio were used in Ref.~\cite{Cholis:2013lwa} to constrain $K_{B} < 10~(13, 16)$ at the 95$\%$ (99$\%$, 99.9$\%$) confidence level.

Starting with the heaviest isotopes, we calculate the spectrum of all secondaries
down to positrons in each SNR,
and then average over the Galactic Disk, assuming a rate of three SNRs per century (see~\cite{Cholis:2013lwa}).
The injected spectrum of CRs in the ISM, after integrating over the volume of the SNR is:
\begin{equation}
N_{i}(E) = 16 \pi^2 \int^{v^{+} \tau^{SN}}_{0} dx \, p^{2} f^{+}_{i}(x, p)  \, (v^{+} \tau^{SN} - x)^2.
\label{eq:SNR_InjSpect}
\end{equation}
We take $\tau^{SN} = 2\times10^{4}$ yr, $v^{+} = 1.25\times10^{7}$ cm s$^{-1}$ and $f^{+}_{i}$ is 
the phase space density of species $i$ down-stream.

%
%
Treating $K_{B}$ as a free parameter, we calculate the spectrum of accelerated secondary antiprotons and protons and compare this result to the $\bar{p}/p$ ratio measured by \textit{AMS-02}. The contribution from accelerated antiprotons is insignificant at low energies, but can increase the $\bar{p}/p$ ratio significantly at energies above $\sim$\,10\,--\,100 GeV. After accounting for the uncertainties described above, we identify a statistical preference for stochastic acceleration.  In Table~\ref{tab:fitStochAccBestFit}, we provide, for each propagation model, the best-fit value of $K_{B}$, along with the 95$\%$ confidence interval for this quantity (corresponding to $\Delta \chi^{2}=2.71$). Even the lower limits on $K_{B}$ are consistently positive, and the fit improves at a level of $\Delta \chi^{2} \simeq$\,10\,--\,21 when accelerated secondaries are included, corresponding to a statistical preference of 3.2\,--\,4.6$\sigma$ \footnote{In the fits we have added in quadrature the reported statistical and systematic errors. At the highest energies the magnet spectrometer resolution and elastic scatterings of protons inside the detector might lead to charge confusion.}. 

\begin{figure}
\hspace{-0.15in}
\includegraphics[width=3.55in,angle=0]{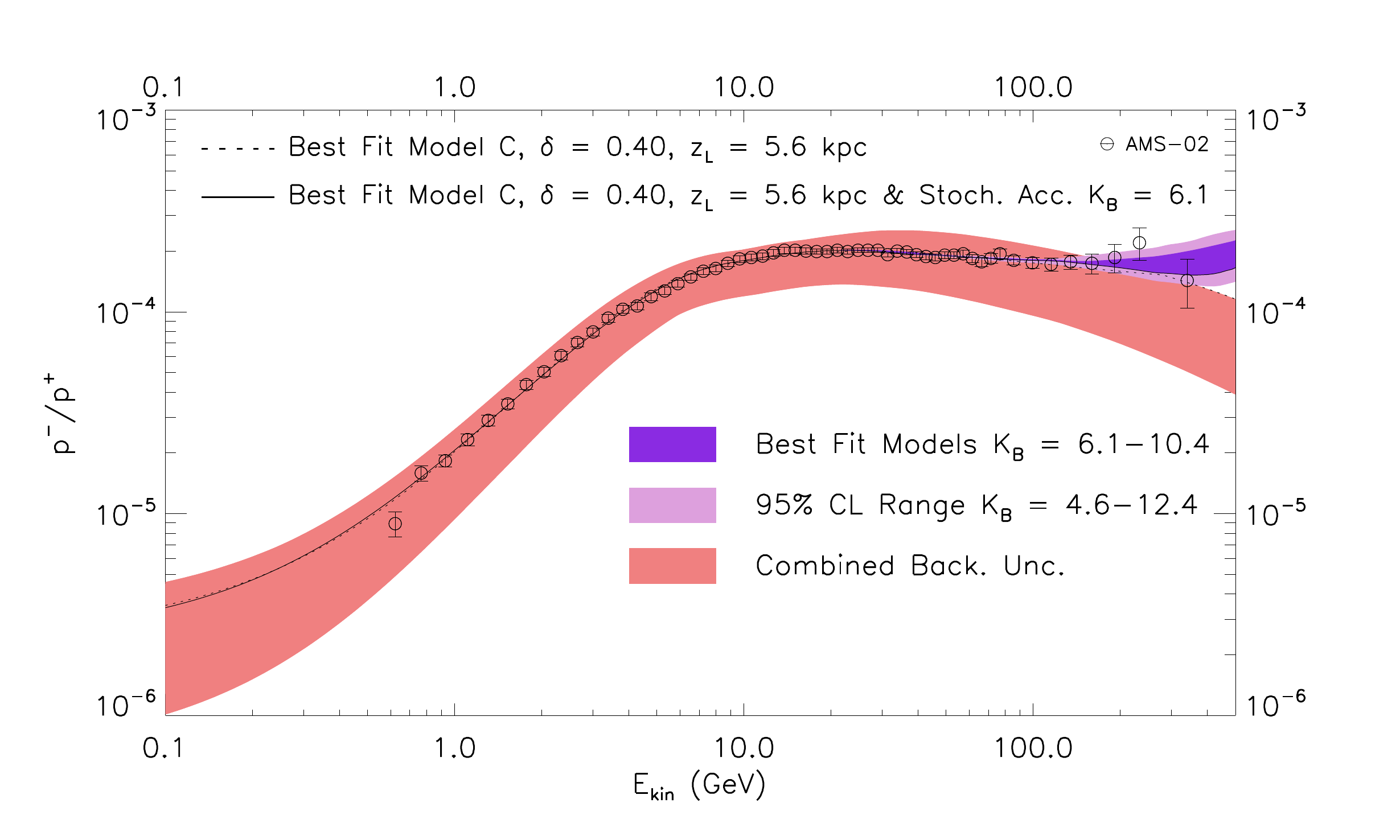}
\vskip -0.1in
\caption{The impact of stochastically accelerated secondaries on the $\bar{p}/p$ spectrum. Accounting for all uncertainties, the best-fit spectra with (without) accelerated secondaries are shown by the solid (dotted) lines. Allowing for accelerated secondaries improves the fit to the $\bar{p}/p$ spectrum.
For the combined $\bar{p}/p$ spectrum, the best fit and 95$\%$ confidence intervals are shown as dark and light purple bands, respectively.}
\label{fig:SASbarp}
\end{figure}

In Fig.~\ref{fig:SASbarp}, we show the impact of accelerated secondary antiprotons on the $\bar{p}/p$ spectrum. The best-fit model is propagation model C with $K_{B} = 6.1$. Given the uncertainties associated with this calculation we also provide a best-fit range (dark purple band) which covers $K_{B}=$ 6.1\,--\,10.4, bracketing the values obtained for the three propagation models considered in this study (see Table~\ref{tab:fitStochAccBestFit}). We also show a 95$\%$ confidence band (light purple) corresponding to $K_{B}=$ 4.6\,--\,12.4. This suggests that on average inside SNRs $B/ \delta B$ is only a factor of few above 1 (between $\simeq 2$ and 3.5) and is in agreement with constraints on ISM CR acceleration \cite{1980ApJ...241.1195C}.
We note that these ranges are consistent with the B/C ratio upper limits of \cite{Cholis:2013lwa}, ($<10$, $13$ at 95, 99$\%$ CL), especially given that the efficiency of SNRs for acceleration of CR secondaries may vary between different environments. 
  
\begin{figure}
\hspace{-0.15in}
\includegraphics[width=3.55in,angle=0]{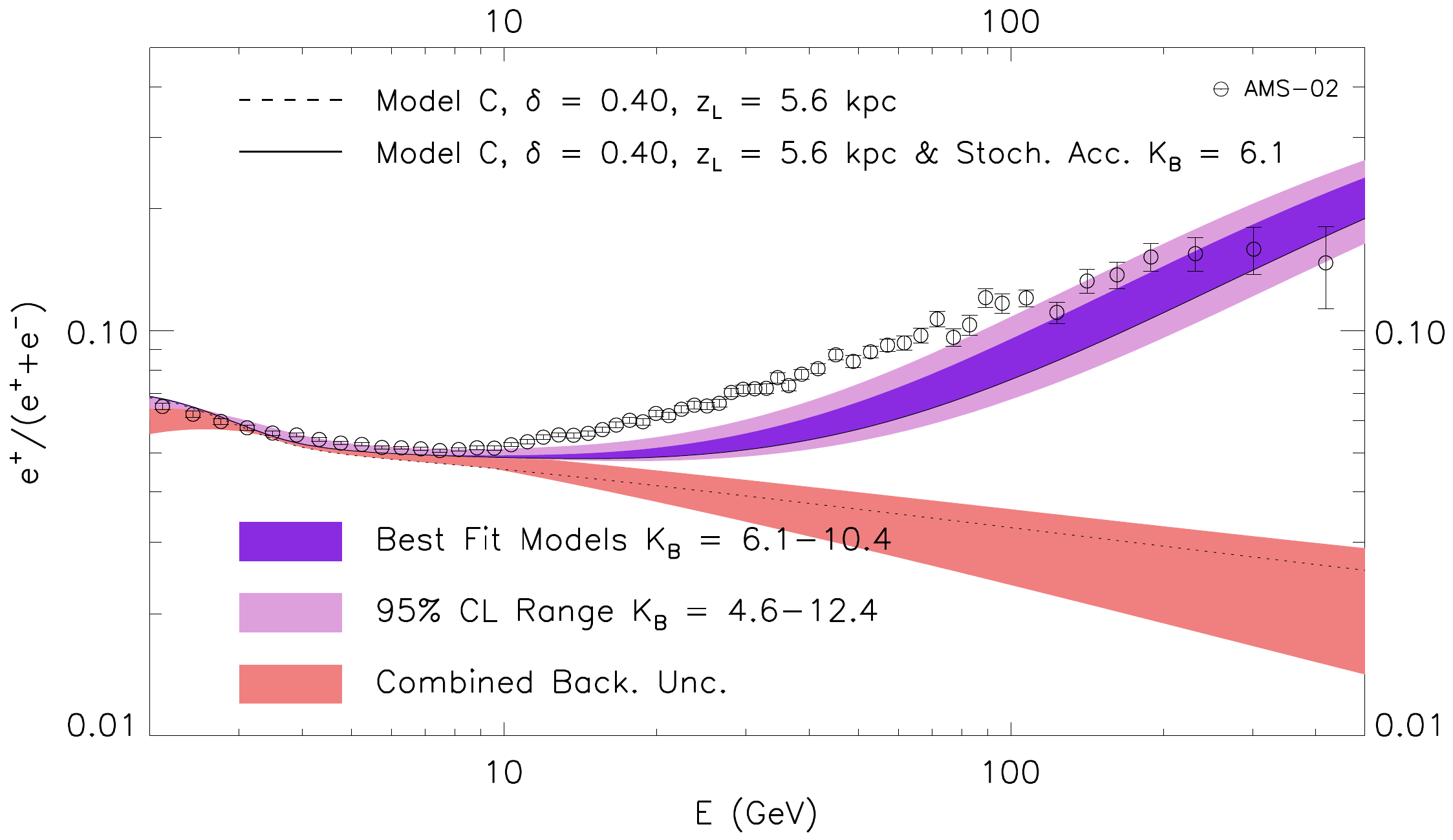}
\vskip -0.1in
\caption{The impact of stochastically accelerated secondaries on the positron fraction measured by \textit{AMS-02}, for the same range of $K_B$ that is required to explain the rising $\bar{p}/p$ ratio of Fig.~\ref{fig:SASbarp}. The measured $\bar{p}/p$ spectrum implies that positrons produced and accelerated in SNRs could account for a significant fraction of the positron excess.}
\label{fig:PosFrac}
\end{figure}

In Fig.~\ref{fig:PosFrac}, we illustrate the impact of accelerated secondaries on the positron fraction, showing the result predicted without the acceleration of secondaries (red band) and including accelerated secondary positrons, using the same range of $K_B$ as shown in Fig.~\ref{fig:SASbarp} (purple bands). We do not include contributions from primary positron sources, such as dark matter or pulsars. The shaded bands account for the combined uncertainties associated with the CR propagation and solar modulation parameters, as well as the local $e^{\pm}$ energy loss rate. For the range of $K_B$ values required to explain the rising $\bar{p}/p$ measured by \textit{AMS-02} \cite{Aguilar:2016kjl}, we predict that accelerated secondary positrons will also account for a significant fraction of the positron excess.

Although we have treated $K_B$ as a simple parameter in this study, this quantity may vary with rigidity. CR diffusion results from particles scattering with random magnetohydrodynamic waves and discontinuities, and thus depends on the spectrum of underlying magnetic perturbations. As such, scattering is only efficient for perturbations on length scales comparable to the Larmor radius of the particle. The spectrum of magnetic perturbations found in SNR environments and future \textit{AMS-02} data will determine the rigidity dependence of $K_B$.


In this paper, we have used the CR $\bar{p}/p$ spectrum, as presented by the \textit{AMS-02} Collaboration, to test scenarios where CR secondaries are produced and accelerated within individual SNRs. The $\bar{p}/p$ spectrum \cite{Aguilar:2016kjl} exhibits a clear rise at energies above 100 GeV. We show that this feature cannot be accounted for by conventional CR sources, even after accounting for the uncertainties pertaining to their injection and propagation through the ISM, the antiproton production cross section, and the effects of solar modulation. Instead, we find that the observed rise is consistent with a contribution of antiprotons that are produced as secondaries and then further accelerated within SNRs. We quantify the range of parameters that can produce this observation, and note that for our best fit models, the acceleration of secondary positrons should contribute substantially to the CR positron flux, potentially accounting for a significant fraction of the observed positron excess.

\bigskip                  
                  
IC acknowledges support from NASA Grant NNX15AB18G and from the Simons Foundation.
DH is supported by the US Department of Energy under contract DE-FG02-13ER41958. Fermilab is operated by Fermi Research Alliance, LLC, under Contract No. DE- AC02-07CH11359 with the US Department of Energy. TL acknowledges support from NSF Grant PHY-1404311. 
FERMILAB-PUB-17-010-A, data from the Advanced Composition Explorer (ACE) Science Center~\cite{ACESite} 
and the Wilcox Solar Observatory obtrained in~\cite{WSOSite}, courtesy of J. T. Hoeksema, were used in this study. The Wilcox Solar Observatory is currently supported by NASA.

\bibliography{AMS02Antiprotons}
\bibliographystyle{apsrev}
\end{document}